\documentclass[11pt]{article}
\usepackage{amsmath}
\usepackage{amssymb}
\usepackage{graphicx}
\usepackage[numbers]{natbib}

\begin{document}

\setcounter{figure}{0}
\setcounter{table}{0}
\setcounter{footnote}{0}
\setcounter{equation}{0}

\vspace*{1.2cm}

\noindent {\large \bf NEW AND UPDATED LONG-PERIODIC TERMS IN HARMONIC\\
 DEVELOPMENT OF THE EARTH TIDE-GENERATING POTENTIAL}
\vspace*{1cm}

\noindent\hspace*{3.5cm} Sergey M. Kudryavtsev$^{a,*}$, Rodolfo G. Cionco$^{b}$
\vspace*{0.2cm}\\
\noindent\hspace*{0.5cm} $^{a}$Sternberg Astronomical Institute, Lomonosov Moscow State University\\
\noindent\hspace*{0.7cm} Universitetsky Pr., 13, Moscow, 119234, Russia 
\vspace*{0.1cm}\\
\noindent\hspace*{0.5cm} $^{b}$CONICET and Environmental Research Group  of the National University of Technology\\
\noindent\hspace*{0.7cm} Col\'on 332, San Nicol\'as, Buenos Aires, 2900, Argentina
\vspace*{0.1cm}\\
\noindent\hspace*{0.6cm} $^{*}$Corresponding author. E-mail: ksm@sai.msu.ru\\

\vspace*{1.3cm}

\noindent {\large \bf Abstract.} We present a new harmonic development of the long-periodic band of the Earth tide-generating potential (TGP). It updates the corresponding part of the previous TGP expansion, KSM03 (Kudryavtsev, J. Geodesy, 77:829, 2004), and includes 38 terms of period longer than $\simeq$18 years (yr) and amplitude not less than $10^{-8}~{\rm m}^2\,{\rm s}^{-2}$. 
The  development is made through a modified spectral analysis of the TGP numerical values tabulated over more than 30,000 yr (13,200~BC--17,191~AD). The latest JPL NASA's long-term numerical ephemeris DE441 (Park et al., Astron. J., 161:105, 2021) is used as the source of the Moon, the Sun and major planets coordinates. 
For comparison, the KSM03 series were done on the basis of an older DE406 ephemeris (Standish, JPL IOM 312.F, 1998) and over a shorter time interval of 2000 yr (1000--3000). 

As a result of using an extended time span several new long-periodic waves in the Earth TGP are found and most of other terms are updated. In particular, a relatively large term of amplitude of $3\times 10^{-5}~{\rm m}^2\,{\rm s}^{-2}$ and period of $\simeq$7.4~kyr is revealed.
Several new waves of period close to 18.61 yr (the period of the lunar nodal cycle, LNC) are separated from the main LNC term.
The effect of the general precession in longitude (of $\simeq$25.7~kyr period) on the Earth TGP for the first time is evaluated. As a result, a number of updated TGP terms include the precession rate in their arguments.

A new catalogue of the long-periodic terms in the Earth TGP spectrum in both standard HW95 and KSM03 format is released. 
 
\vspace*{0.5cm}


\vspace*{1.5cm}

\noindent {\Large \bf 1. Introduction}\label{1}

\smallskip

The concept of the tide-generating potential (TGP) in investigating the Earth tides was suggested by Laplace \cite{Laplace1790} in 1790. In his subsequent studies he elaborated the TGP formalism and made the first decompositions of the tidal potential to harmonic series. These series were then used and further extended in the classical studies by Thomson \cite{Thomson1869} in 1868, Ferrel \cite{Ferrel1874} in 1874, Darwin \cite{Darwin1883} in 1883, and Doodson \cite{Doodson1921} in 1921.

During the last decades a number of advanced Earth TGP harmonic expansions were done by 
Cartwright \& Tayler \cite{Cartwright1971}, % in 1971, 
Cartwright \& Edden \cite{Cartwright1973}, % in 1973, 
B\"ullesfeld \cite{Bullesfeld1985}, % in 1985, 
Xi \cite{Xi1987, Xi1989}, % in 1987 and 1989, 
Tamura  \cite{Tamura1987,Tamura1995}, % in 1987 and 1995, 
Roosbeek \cite{Roosbeek1996} (called as RATGP95), % in 1996, 
Hartmann \& Wenzel \cite{Hartmann1994,Hartmann1995,Hartmann1996} (HW95), % in 1996, 
Kudryavtsev \cite{Kudryavtsev2004,Kudryavtsev2007a} (KSM03). % in 2004 and 2007. 
Several new TGP terms due to the Earth's triaxiality were recently suggested in \cite{Kudryavtsev2025}.
 
Most of the early expansions were built by purely analytical methods but the latest Earth TGP developments (HW95, KSM03) were made with use of some numerical methods of the spectral analysis. In the latter case one processes the TGP values pre-calculated with a small sampling step over a long interval of time. This approach allows one to employ the most accurate numerical ephemerides of the tide-generating bodies (the Moon, the Sun and major planets). Also the precision algorithms for performing the necessary transformations of the bodies coordinates (e.g., those recommended by the IERS Conventions \cite{IERS2003,IERS2010}) can be used. 

However, a common issue of all known methods of the spectral analysis is the problem of ``close frequencies'', but it can be solved (or diminished) by increasing the time interval where the TGP values are processed. In particular, the latest KSM03 development is made by means of the spectral analysis of the TGP values calculated over 2000 years (yr), where the JPL's planetary and lunar ephemerides DE405/406 \cite{Standish1998} of 1998 were used as the source. 
Since that time several new long-term numerical ephemerides of the Moon and major planets were released in various organizations, e.g., the latest JPL's ephemeris, DE441  \cite{Park2021} of 2021, covers the time span of more than 30,000 yr (from -13,200 to +17,191).
Using such the long time interval in developments of the Earth TGP should allow one to better separate different terms of close frequencies.
In particular, separating the terms of periods close to 18.61 yr (the period of the lunar nodal cycle, LNC) in the Earth TGP would be of special interest; the main term of the LNC period has the largest amplitude among all TGP expansion terms that have a period of one year or more. 

In addition, new terms of very large periods (of hundreds and thousands of years) can potentially be revealed when developing the Earth TGP over such an extended time interval. 

Therefore, the main aim of the present study is to make a new development of the long-periodic part of the Earth TGP on the basis of  modern and essentially extended ephemeris data.  
We shall also use the up-to-date values for the planetary/lunar masses and the new precession theory \cite{Vondrak2011}, specially developed for its application over very long time intervals (like tens and hundreds of thousands of years). So, the parameters of the already known terms of the Earth TGP development series should be improved as well.

\vspace*{1cm}

\noindent {\Large \bf 2. Method and data}\label{2}

\smallskip

\begin{figure}
\centering
	\includegraphics[width=.9\columnwidth]{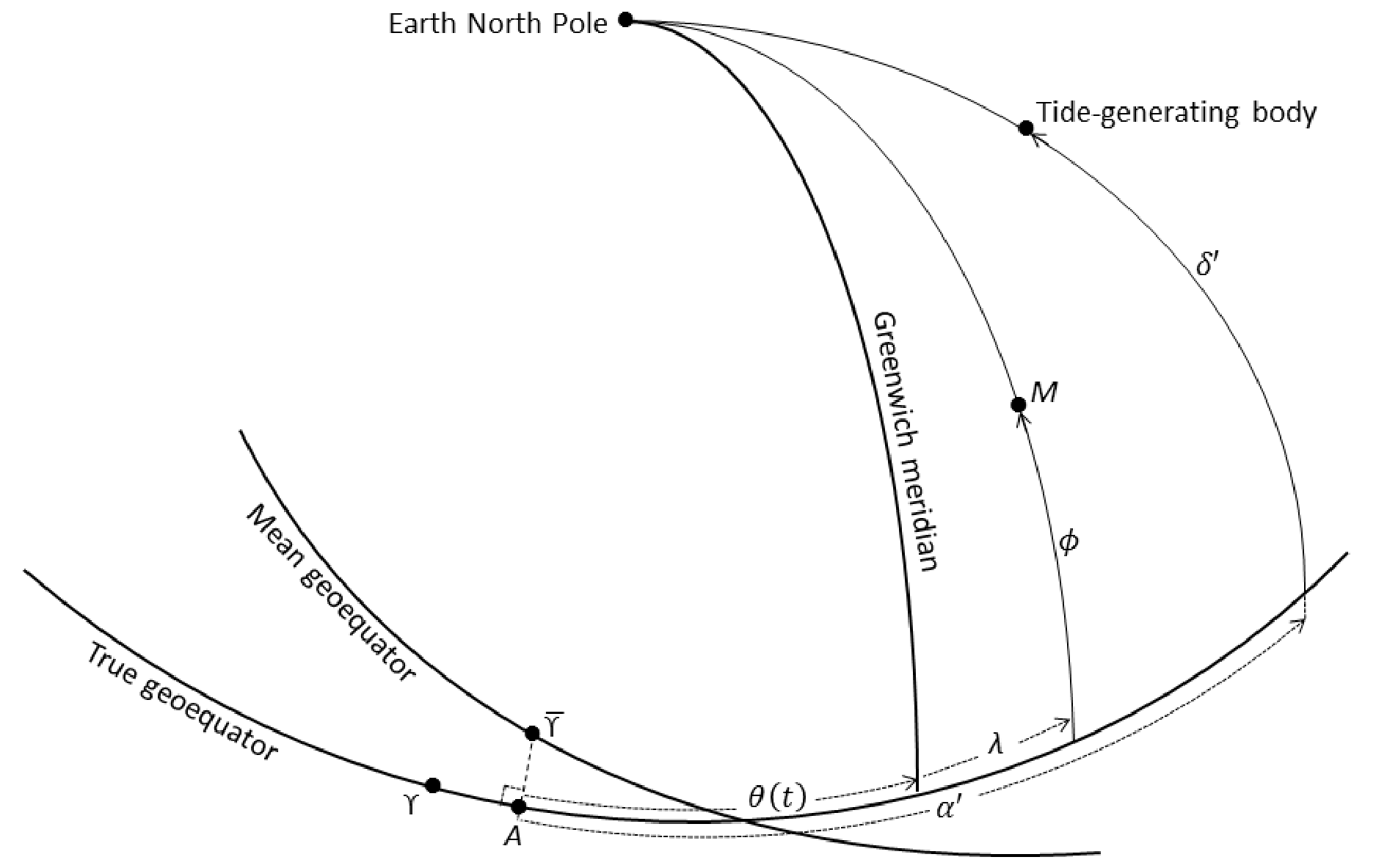} 
\caption{The coordinate system used when developing the Earth TGP at site $M$.} 

\label{Fig1}
\end{figure}

Let $M(r,\phi,\lambda)$ be an arbitrary site on the Earth's surface (or above it) 
where $r$ is the geocentric distance of $M$, and $\phi$, $\lambda$ are, respectively, the site geographic latitude and 
longitude reckoned from the Greenwich meridian to the East (Fig.~\ref{Fig1}).
The value of the Earth TGP at $M$ at epoch $t$ can be represented as follows \cite{Kudryavtsev2004} 
\begin{equation}
V(t) = \sum_{n=1}^{\infty} \left(\frac{r}{\rm{R_{E}}}\right)^n \sum_{m=0}^n {\bar P}_{nm}(\sin \phi) 
% \nonumber \\
% & \times & 
\left[
%     \begin{array}{l}
      C_{nm}(t) \cos m (\lambda+\theta(t))  \\ % + S_{nm}(t)\sin m (\lambda+\theta(t)) \right],
      + S_{nm}(t)\sin m (\lambda+\theta(t))  
%     \end{array}   
     \right],
\label{eq1}
\end{equation}

\noindent where $\rm{R_{E}}$ is the Earth mean equatorial radius; ${\bar P}_{nm}$ are the normalized associated Legendre functions of degree $n$ and order $m$ 
related to the non-normalized ones, $P_{nm}$, as  
\begin{eqnarray}
  {\bar P}_{nm}=\sqrt{\frac{\delta_m (2n+1)(n-m)!}{(n+m)!}} \, P_{nm},   \,\,\,
 \nonumber \\
   \delta_m=\left\{
    \begin{array}{l}
     1, \quad {\rm if} \: m=0 \\
     2, \quad {\rm if} \: m \ne 0  
   \end{array}
   ,\right.   
	\label{eq2}
\end{eqnarray}

\noindent $\theta (t)$ is the Greenwich Mean Sidereal Time at epoch $t$ \cite{Capitaine2005}, it is reckoned along the true geoequator from point $A$ - the projection of the 
mean equinox $\bar {\Upsilon}$ on the true geoequator of epoch $t$ (Fig.~\ref{Fig1}); $C_{nm}(t)$ and $S_{nm}(t)$ are some time-variable coefficients depending on 
the mass and instantaneous geocentric distance $r'$, right ascension $\alpha'$ and declination $\delta'$ of an external tide-generating body (the Moon, the Sun, major planets). 
The exact view of $C_{nm}(t)$ and $S_{nm}(t)$ is 
given in \cite{Kudryavtsev2004}, and the task of the Earth TGP development mostly amounts to representing these 
%time-variable 
coefficients by harmonic series.

From Eq.~(\ref{eq1}) one sees that if $m \neq 0$ then the corresponding terms of the TGP include a fast changing variable $\theta (t)$ so that the periods of such 
tidal waves are of around one day ($m=1$) or less ($m>1$). Therefore, as far as in the present study we look for the long-periodic part of the Earth TGP we further consider the terms of the order $m=0$ only, and the corresponding part of the TGP, designated as $V_0$, is as follows
%Eq.~(\ref{eq1}) is simplified as follows
\begin{equation}
V_0(t) = \sum_{n=1}^{\infty} \left(\frac{r}{\rm{R_{E}}}\right)^n  {\bar P}_n (\sin \phi) C_{n}(t) ,
\label{eq3}
\end{equation} 

\noindent where ${\bar P}_n$ is the normalized Legendre polynomial of degree $n$ and $C_{n}(t) \equiv  C_{n0}(t)$. 

In order to expand $C_{n}(t)$ ($n=1,2,...$) to harmonic series, we first tabulate numerical values of the coefficients over the total time interval covered by the DE441 ephemerides (30,000+ yr) with a small sampling step of one day. 
Then for every coefficient we approximate the sampled values by a harmonic series with help of a modified method of spectral analysis \cite{Kudryavtsev2004,Kudryavtsev2007b}. 
A feature of the method is that it evaluates the spectrum at arguments that are polynomials of high orders of time. The amplitudes of the series terms are represented by time polynomials as well  
(unlike the classical Fourier analysis does, where both amplitudes and frequencies of the series terms are assumed to be constants). 
Earlier, the method proved its applicability over the same time interval of 30,000+ yr to harmonic development of several other important functions and potentials of celestial mechanics 
\cite{Kudryavtsev2016,Kudryavtsev2017,Cionco2021,Cionco2023,Kudryavtsev2024,Kudryavtsev2025}.

As arguments of the series terms, in the present study we use various linear combinations of integer multipliers
of thirteen fundamental variables of time:  
\begin{itemize}
\item  $l_1(t) =$~the mean orbital longitude of Mercury, hereafter called as $Me$, the current period is 88.0 days (d);  
\item  $l_2(t) =$~the mean orbital longitude of Venus, $V$, period 224.7 d;
\item  $l_3(t) =$~the mean orbital longitude of Mars, $Ma$, period 1.88 yr;
\item  $l_4(t) =$~the mean orbital longitude of Jupiter, $J$, period 11.86 yr;
\item  $l_5(t) =$~the mean orbital longitude of Saturn, $S$, period 29.42 yr;
\item  $l_6(t) =$~the mean orbital longitude of Uranus, $U$, period 83.75 yr;
\item  $l_7(t) =$~the mean orbital longitude of Neptune, $N$, period 163.72 yr;
\item  $l_8(t) =$~the mean orbital longitude of the Moon, $L_M$, period 27.32 d;
\item  $l_9(t) =$~the mean orbital longitude of the Sun, $L_S$, period 365.24 d;
\item  $l_{10}(t) =$~the mean orbital longitude of the lunar perigee, $P_M$, period 8.85 yr;
\item  $l_{11}(t) =$~the negative mean orbital longitude of the lunar ascending node, $N^{'}$, period 18.61 yr; 
\item  $l_{12}(t) =$~the mean orbital longitude of the solar perigee, $P_S$, period $\approx$ 20,9 kyr;
\item  $l_{13}(t) =$~the general precession in longitude, $p_A$, period $\approx$ 25,7 kyr.
\end{itemize}
Mean orbital longitudes of the major planets were taken from \cite{Simon1994} and those of the Moon were derived from \cite{Chapront2002}. 
All orbital longitudes are referred to the mean equinox and ecliptic of date and were
obtained with use of the time polynomial part of the general precession in longitude \cite{Vondrak2011}. 
As a result, the employed expressions for the fundamental variables are given by time polynomials of up to the sixth order. 
It allows one to use them in the arguments of harmonic series for representing the tabulated data over long time intervals, up to thousands and tens of thousands years. 
It is important for the successive spectral analysis of the data because the longer data series is used, the larger periodicities in the TGP can be identified and the better separation of the series terms of close frequencies can be achieved.

We shall note, the general precession in the longitude was not explicitly included in the terms arguments of the Earth TGP harmonic series developed to now. However, it could be done in the present study due to the large time span of the source data that exceeds the period of the precession. So, the direct effect of that new fundamental variable on the Earth TGP harmonic development can be taken into account and investigated. 

When calculating the terms arguments as different linear combinations of the fundamental variables, we varied the corresponding integer multipliers up to $\pm20$. 
Therefore, in total around 9.4 million arguments have been considered during the modified spectral analysis of the tabulated data for every $C_{n}(t)$ ($n=~1,2,\dots$) coefficient. 
A general form of the argument of the $i^{th}$ term in the development series for any $C(t)$ coefficient is as follows (hereafter for simplicity's sake we omit the subscript $n$)
\begin{equation}
\omega_{i}(t) = \sum_{j=1}^{13} k_{ij} l'_j(t)=\nu_{{1}i}t+\nu_{{2}i}t^2+\dots+\nu_{{6}i}t^6, 
\label{eq4}
\end{equation} 

\noindent where $\omega_{i}(t)$ is the argument, $l'_j(t)$ $(j=1,2,...,13)$ is the time-dependent part of the $l_j(t)$ fundamental variable (not including the constant term of $l_j(t)$), $k_{{ij}}$ is a set of some integer multipliers, and $\nu_{{h}i}$ ($h=~1,2,\dots,6$) is a numerical value for the $h$-order frequency in the $i^{th}$ term argument calculated on the basis of the relevant values for $k_{ij}$ and $l'_j(t)$. 

At every argument we evaluated the amplitude of the corresponding term that, in a general case, was
represented by a sixth-order polynomial of time. (However, for tidal waves of very long periods 
only some low-order terms of their polynomial amplitudes could be determined, 
otherwise the time interval of the tabulated data should be even longer.) 
So, every $C(t)$ coefficient was represented by a series of the following form
\begin{equation} 
      C(t) = \sum_{i}\left[\left(A^{\rm c}_{{0}i}+A^{\rm c}_{{1}i}t+\dots+A^{\rm c}_{{6}i}t^6\right) \cos\omega_{i}(t) \right.
% \nonumber \\
            + \left. \left(A^{\rm s}_{{0}i}+A^{\rm s}_{{1}i}t+\dots+A^{\rm s}_{{6}i}t^6\right) \sin\omega_{i}(t)\right],
\label{eq5}
\end{equation}

\noindent where amplitudes $A^{\rm c}_{{0}i}, A^{\rm c}_{{1}i}, \dots, A^{\rm c}_{{6}i}, A^{\rm s}_{{0}i}, A^{\rm s}_{{1}i}, \dots, A^{\rm s}_{{6}i}$, and frequencies $\nu_{{1}i}, \nu_{{2}i}, \dots, \nu_{{6}i}$ of the argument $\omega_{i}(t)$ (see  Eq.~(\ref{eq4})) are some numerical constants determined in the course of our spectral analysis of the pre-calculated values for the Earth TGP. 

\vspace*{1cm}

\noindent {\Large \bf 3. Results}\label{3}

\smallskip

Following the procedure described above, we created an updated catalogue of the long-periodic part of the Earth TGP, called KC25. It includes 38 terms of period longer than $\simeq$18 yr and amplitude not less than $10^{-8}~{\rm m}^2\,{\rm s}^{-2}$ (such the threshold for amplitudes was also selected in the previous tidal potential catalogues \cite{Hartmann1995,Hartmann1996,Kudryavtsev2004,Kudryavtsev2007a}). 
The new catalogue is available in {\tt http://lnfm1.sai.msu.ru/neb/ksm/tgp/kc25\_format-ksm03.zip}. It contains all numerical coefficients of the sixth-order time polynomials representing both arguments (Eq.~(\ref{eq4})) and amplitudes (Eq.~(\ref{eq5})) of the KC25 expansion along with the corresponding set of the integer multipliers of the involved fundamental variables. The data are given in a modified KSM03 format \cite{Kudryavtsev2004}, it is described in {\tt readme.pdf} file included in the archive.

The new development of the long-periodic part of the Earth TGP is also transformed to the standard HW95 format \cite{Hartmann1995,Hartmann1996} and available in {\tt http://lnfm1.sai.msu.ru/neb/ksm/tgp/ kc25\_format-hw95.dat.} That format by us was generalized as follows. We added four new fields for the numerical coefficients at the terms of the second- and third-order of time in the polynomials for amplitudes and three more fields for the integer multipliers of additional fundamental variables (the orbital longitudes of Uranus and Neptune and the general precession in longitude).

In the vicinity of the initial epoch $t_0$ of the development (chosen as J2000.0) the complete expansion series defined by Eqs.~(\ref{eq4})-(\ref{eq5}) can be simplified as

\begin{equation} 
      C(t) \approx \sum_i A_{0i} \cos\left(\nu_{1i} \ (t-t_0) + \varphi_{i}\right),
\label{eq6}
\end{equation}

\noindent where  $A_{0i}$ is the constant part of the amplitude of the $i^{th}$ term at the initial epoch $t_0$ 
\begin{equation} 
   A_{0i}=\sqrt{{A_{0i}^{\rm c}}^2+{A_{0i}^{\rm s}}^2},
\label{eq7}
\end{equation} 

\noindent and $\nu_{1i}$, $\varphi_{i}$ are, respectively, the main frequency of the argument $\omega_{i}$ and the corresponding phase of the $i^{th}$ term at $t_0$.  
Both amplitudes, frequencies and phases of the series terms in Eq.~(\ref{eq6}) can be recalculated at an arbitrary initial epoch by using the complete set of numerical coefficients from Eqs.~(\ref{eq4})-(\ref{eq5}).  

Table~\ref{table1} gives the main characteristics of the new and updated long-periodic and constant terms in the Earth TGP harmonic series represented by Eq.~(\ref{eq6}) at the initial epoch J2000.0. It merges the data for the $C_{n}(t)$ coefficients of $n \le 4$; the amplitudes of  long-periodic terms in the TGP series for the coefficients of a higher degree $n$ were proven to be below the selected threshold level. Additionally, for every tidal potential wave included in Table~\ref{table1}, the corresponding period $T$ and the integer multipliers related to the involved fundamental variables are given. The arguments of the long-periodic terms in the Earth TGP revealed in the present study do not include the mean orbital longitudes of Mercury, Uranus and Neptune, so the corresponding integer multipliers are zero and we do not show them in Table~\ref{table1}. The terms are ranked by decreasing amplitude $A_0$. 

\begin{table}%[]
\caption{Long-periodic and constant terms of the Earth TGP harmonic development ranked by decreasing amplitude
(all parameters are given at the epoch J2000.0).} 
\label{table1}      % is used to refer this table in the text
%\centering                          % used for centering table
\begin{tabular}{rrrrrrrrrrrrrrrrrrrr}     
%\begin{tabular}{llllllllllllllllllll}    
%\begin{tabular*}{\tblwidth}% {@{}LL@{}}
%\toprule
\hline
 Rank & $n$  & $A_{0}\times 10^8$ & $\nu_{1}$ & $\varphi$ & $T$ &  \multicolumn{10}{l}{Integer multipliers for fundamental variables}  \\    % table heading 
% \cline{7-16} \\
 &  & [m$^2$s$^{-2}$] &  [rad\,yr$^{-1}$] & [rad] &  [yr] &  $V$ & $Ma$ & $J$ & $S$ & $L_M$ & $L_S$ & $P_M$ & $N^{'}$ & $P_S$ & $p_A$  \\
%\midrule
\hline
   1   &   2 &  86957109.1 &      0.0000 &  3.1416 &     Infinity &      0 &   0 &   0 &   0 &      0 &   0 &   0 &   0 &   0 &   0 \\
  2   &   2 &   7719972.3 &      0.3376 &  4.1003 &      18.613 &      0 &   0 &   0 &   0 &      0 &   0 &   0 &   1 &   0 &   0 \\
  3   &   4 &      3753.8 &      0.0000 &  0.0000 &     Infinity &      0 &   0 &   0 &   0 &      0 &   0 &   0 &   0 &   0 &   0 \\
  4   &   2 &      3054.8 &      0.0009 &  1.0001 &    7426.248 &      0 &   0 &   0 &   0 &      0 &   0 &   0 &   0 &   2 &   1 \\
  5   &   4 &      2853.7 &      0.3376 &  0.9591 &      18.613 &      0 &   0 &   0 &   0 &      0 &   0 &   0 &   1 &   0 &   0 \\
  6   &   2 &      1755.9 &      0.3370 &  5.8019 &      18.646 &      0 &   0 &   0 &   0 &      0 &   0 &   0 &   1 &  -2 &   0 \\
  7   &   2 &      1506.9 &      0.3382 &  5.5562 &      18.580 &      0 &   0 &   0 &   0 &      0 &   0 &   0 &   1 &   2 &   0 \\
  8   &   3 &        65.9 &      0.0003 &  0.2327 &   20888.805 &      0 &   0 &   0 &   0 &      0 &   0 &   0 &   0 &   1 &   0 \\
  9   &   2 &        40.3 &      0.3364 &  3.0122 &      18.680 &      0 &   0 &   0 &   0 &      0 &   0 &   0 &   1 &  -4 &   0 \\
 10   &   3 &        28.5 &      0.0350 &  1.1073 &     179.325 &      0 &   0 &   0 &   0 &      0 &   0 &   1 &  -2 &   0 &   0 \\
 11   &   2 &        14.4 &      0.3342 &  3.1736 &      18.803 &      1 &   0 &   0 &   0 &      1 & -15 &   1 &  -1 &   0 &   3 \\
 12   &   2 &        11.9 &      0.3410 &  1.8642 &      18.427 &     -1 &   0 &   0 &   0 &     -1 &  15 &  -1 &   3 &   0 &  -3 \\
 13   &   2 &        10.4 &      0.1692 &  1.8689 &      37.146 &      0 &   0 &   3 &   0 &      0 &   0 &  -2 &   0 &  -1 &   0 \\
 14   &   2 &        8.8  &      0.2131 &  0.3682 &      29.491 &      0 &   0 &   0 &   1 &      0 &   0 &   0 &   0 &   0 &  -2 \\
 15   &   2 &        8.4  &      0.0074 &  2.0611 &     853.915 &      0 &   0 &  -2 &   5 &      0 &   0 &   0 &   0 &   0 &  -2 \\
 16   &   3 &        7.1  &      0.3379 &  4.3205 &      18.596 &      0 &   0 &   0 &   0 &      0 &   0 &   0 &   1 &   1 &   0 \\
 17   &   2 &        6.2  &      0.0260 &  1.9323 &     241.864 &      8 &   0 &   0 &   0 &      0 & -13 &   0 &   0 &   3 &   0 \\
 18   &   2 &        5.1  &      0.1263 &  4.6970 &      49.768 &      5 &   0 &   0 &   0 &      0 &  -8 &   0 &  -2 &   1 &   0 \\
 19   &   2 &        4.4  &      0.1921 &  1.1767 &      32.714 &      0 &   0 &   1 &   0 &      0 &   0 &   0 &  -1 &  -1 &   0 \\
 20   &   2 &        4.4  &      0.0605 &  4.7654 &     103.815 &      0 &   2 &   0 &   0 &      0 &  -1 &   0 &  -1 &  -1 &   0 \\
 21   &   2 &        3.7  &      0.3309 &  3.8581 &      18.990 &      8 &   0 &   0 &   0 &     -2 &  14 &   0 &  -4 &   2 &   0 \\
 22   &   2 &        3.7  &      0.3444 &  1.1267 &      18.246 &      0 &   0 & -10 &   0 &      0 &   1 &   1 &  -4 &   0 &   2 \\
 23   &   2 &        3.3  &      0.3144 &  2.0372 &      19.988 &      0 &   0 &   2 &   0 &      0 &   0 &  -2 &   2 &  -1 &   0 \\
 24   &   2 &        3.3  &      0.3388 &  2.8990 &      18.547 &      0 &   0 &   0 &   0 &      0 &   0 &   0 &   1 &   4 &   0 \\
 25   &   2 &        3.2  &      0.3113 &  3.5282 &      20.184 &      0 &  -2 &   0 &   0 &      0 &   1 &   1 &   0 &   0 &  -2 \\
 26   &   3 &        2.9  &      0.3497 &  4.4596 &      17.968 &      0 &   0 &   2 &   0 &      0 &   0 &  -1 &   0 &   0 &   0 \\
 27   &   3 &        2.2  &      0.3373 &  0.6913 &      18.630 &      0 &   0 &   0 &   0 &      0 &   0 &   0 &   1 &  -1 &   0 \\
 28   &   2 &        2.0  &      0.1684 &  2.2082 &      37.319 &      0 &   0 &  -3 &   0 &      0 &   0 &   2 &   1 &   0 &   1 \\
 29   &   2 &        1.8  &      0.1312 &  2.0170 &      47.898 &     -6 &   0 &   0 &   0 &      0 &  10 &  -2 &   0 &   0 &  -2 \\
 30   &   3 &        1.6  &      0.0658 &  2.4890 &      95.440 &     -3 &   0 &   0 &   0 &      0 &   5 &  -1 &   0 &   0 &   0 \\
 31   &   2 &        1.6  &      0.1560 &  5.9303 &      40.278 &      0 &   0 &  -5 &   0 &      0 &   0 &   3 &   2 &   0 &   0 \\
 32   &   2 &        1.5  &      0.0232 &  4.6503 &     270.650 &      0 &   0 &  -2 &   0 &      0 &   0 &   2 &  -1 &   1 &   0 \\
 33   &   2 &        1.5  &      0.0655 &  5.6360 &      95.878 &     -3 &   0 &   0 &   0 &      0 &   5 &  -1 &   0 &  -1 &   0 \\
 34   &   2 &        1.3  &      0.1802 &  3.9747 &      34.860 &      0 &   0 &  -1 &   0 &      0 &   0 &   1 &   0 &   0 &   0 \\
 35   &   2 &        1.3  &      0.3162 &  4.0575 &      19.874 &      0 &   0 &   1 &  -1 &      0 &   0 &   0 &   0 &   0 &  -1 \\
 36   &   2 &        1.2  &      0.1246 &  4.0188 &      50.445 &      0 &   0 &   0 &   2 &      0 &   0 &   1 &  -3 &   0 &   0 \\
 37   &   2 &        1.0  &      0.2254 &  1.8838 &      27.873 &      0 &  -6 &   0 &   0 &      0 &   3 &   2 &   0 &   0 &   1 \\
 38   &   2 &        1.0  &      0.2720 &  4.7805 &      23.097 &      3 &   0 &   0 &   0 &      0 &  -5 &   1 &   1 &   1 &   0 \\
%\bottomrule
 \hline
\end{tabular}
%\end{tabular*} 
\end{table}

\vspace*{1cm}

\noindent {\Large \bf 4. Discussion}\label{4}

\smallskip

It is well known that the retrograde motion of the lunar orbit nodes causes the largest long-periodic TGP waves (the terms of rank 2 and 5 in Table~\ref{table1}). In a narrow spectral band around the LNC period of 18.61 yr several close periodicities are revealed. For example, there are some terms of degree two with periods of 18.65, 18.58, 18.68 and 18.55~yr (ranks 6, 7, 9, and 24, respectively) as well as similar terms of degree three with periods of 18.60 yr and 18.63 yr (ranks 16 and 27, respectively). 
Those terms are produced by some linear combinations of the negative longitude of the lunar ascending node ($N'$) and the longitude of the solar perigee ($P_S$). 
Terms of similar periodicities that involve the orbital longitudes of major planets are also observed (e.g., rank 11 and 12). 
The identification of such periods, close to that of the LNC term, was possible due to using the large time interval of 30,000+ yr in the development. 

On the multi-decadal scale we have tidal waves with periods of 37.15 yr and 29.49 yr (ranks 13 and 14, respectively), they are caused by the orbital motion of Jupiter and Saturn combined with the motion of the lunar or solar perigee along with the general precession. There are also several smaller terms that have periods of $\simeq$20--50 yr and $\simeq$95--96 yr.
It worths noting that several studies argue in favour of the existence of an $\sim60$ yr periodicity in tidal gauges, e.g., \cite{Pan2021}. %\cite{Ding2021,Pan2021}
 Although the causes of the phenomenon currently remain unclear and are likely of non-astronomical origin 
\cite{Haighetal2019s},
 certain astronomical oscillations were also assumed in this context \cite{Scafetta2014}. However, no tidal potential wave of a periodicity close to 60 yr is detected in the present expansion of the Earth TGP.

On the multi-centennial scale the most important term is one of the third degree of 179.33 yr period (rank 10). 
A new TGP wave of 853.9 yr period (rank 15) in this band is found. It is attributed to the Great Inequality between Jupiter and Saturn affected by the general precession (argument $-2J+5S-2p_{A}$). 

On the multi-millennial scale we see a relatively large wave of $\simeq$~7.4~kyr period (rank 4); it is not available in any previous development of the Earth TGP. As we found, this wave is ``originated'' by the second harmonic of the solar perigee motion acting along with the general precession (argument $2P_S+p_A$). In general, there is an aliasing effect when using the fundamental variables $P_S$ (period $\simeq20.9$~kyr) and $p_A$ (period $\simeq25.7$~kyr) simultaneously. Therefore, we had to use these two variables separately in the most of cases. However, for the term of an $\simeq7.4$~kyr period the simultaneous usage of both variables did lead to the best identification of the wave source. 
It is interesting, in \cite{Datta2025s} a significant spectral power at 7--8~kyr in ``benthic foraminifera'' microorganisms was shown, implying multi-millennial cyclic changes in the Southern Ocean during the late Pleistocene. It is still unexplained, so a future investigation of a possible connection between that phenomenon and the found new tidal wave of $\simeq7.4$~kyr period would be useful.

The present TGP development includes the third-degree term of $\simeq20.9$~kyr period produced by the solar perigee motion (rank 8). It is available in some TGP expansions \cite{Roosbeek1996,Hartmann1995}, but in our previous development \cite{Kudryavtsev2004} that long-periodic wave was absent. Instead, a constant term (or ``permanent tide'') in the $C_{3}(t)$ coefficient was given, it was discussed in  \cite{Makinen2021} then. The present update of the respective part of the TGP confirms our earlier assumption (referred as a personal communication in \cite{Makinen2021}) that in \cite{Kudryavtsev2004} the constant term in $C_{3}(t)$ is just an artefact. It appeared when making the spectral analysis of the TGP values over a time interval of 2000 yr, that is much smaller than the period of the discussed wave. Here the spectral analysis was done over 30,000+ yr and the term of an $\simeq20.9$~kyr period in the $C_{3}(t)$ tidal coefficient was confidently revealed but the ``constant'' term vanished.    

We did not find other TGP waves of a multi-millennial period, in particular, the terms that are solely due to the first- or higher-order harmonic of the general precession. However, they might still be revealed when investigating the Earth TGP over longer time intervals, like hundreds of thousands of  years and more. 

The new KC25 development increases the number of detected long-periodic waves in the Earth TGP. In total, the KC25 series in the HW95 format includes terms of 35 different periods longer than $\simeq$18 yr. For comparison, the KSM03 series given in that format \cite{Kudryavtsev2007a} includes 21 terms of such periods, HW95 \cite{Hartmann1995} - 14, RATGP96 \cite{Roosbeek1996} - 8, and older expansions \cite{Doodson1921,Cartwright1971,Cartwright1973,Tamura1987,Xi1987} - 2 (the constant term and one of the 18.61 yr period). 
Also, the amplitudes and arguments of the long-periodic TGP terms already available in the KSM03 development \cite{Kudryavtsev2004,Kudryavtsev2007a} have been updated.

\vspace*{1cm}

\noindent {\Large \bf 5. Summary and conclusions}\label{5}

\smallskip

\begin{itemize}
\item  A new harmonic development of the long-periodic part of the Earth TGP over 30,000+ yr, KC25, is made. It is based on the latest JPL NASA's  long-term ephemeris of the Moon and major planet DE441 \cite{Park2021}; 

\item The development includes 38 terms of period longer than $\simeq$18 yr and amplitude not less than $10^{-8}~{\rm m}^2\,{\rm s}^{-2}$. New long-periodic waves in the Earth TGP (e.g., those of $\simeq$0.85~kyr and $\simeq$7.4~kyr period) are revealed and parameters of the known terms are updated; 

\item Several new waves of period close to 18.61 yr in the Earth TGP spectrum are found and separated from the main LNC term;

\item For the first time the general precession in longitude is explicitly considered in the TGP terms arguments as an independent fundamental variable. It allows one to get a better expansion of the Earth TGP over long-term time intervals, like tens of thousands years;

\item The KC25 harmonic development of the long-periodic part of the Earth TGP is made available in electronic form in two formats: KSM03 and HW95.

\end{itemize}

\vspace*{2.6cm}

\noindent {\Large \bf CRediT authorship contribution statement}\label{6}

\smallskip

{\bf Sergey M. Kudryavtsev:} Conceptualization, Methodology, Software, Investigation, Writing. {\bf Rodolfo G. Cionco:} Conceptualization, Methodology, Writing.

\vspace*{1cm}

\noindent {\Large \bf Data availability statement}\label{7}

\smallskip

The data obtained in this study are available in electronic form in {\tt http://lnfm1.sai.msu.ru/neb/ksm/ tgp/kc25\_format-ksm03.zip} and {\tt http://lnfm1.sai.msu.ru/neb/ksm/tgp/kc25\_format-hw95.dat}.

\vspace*{1cm}

\noindent {\Large \bf Declaration of competing interest}\label{8}

\smallskip

We declare that we do not have any commercial or associative
interest that represents a conflict of interest in connection with the
work submitted.

\vspace*{1cm}

\noindent {\Large \bf Acknowledgements}\label{9}

\smallskip

The study was conducted under the state assignment of Lomonosov Moscow State University.

\vspace*{1cm}

{

%\leftskip=5mm
%\parindent=-5mm

\smallskip

\bibliographystyle{ieeetr}
\bibliography{references}

}

\end{document}